# Harder, better, faster, stronger: large-scale QM and QM/MM for predictive modeling in enzymes and proteins

Vyshnavi Vennelakanti[1,2], Azadeh Nazemi[1], Rimsha Mehmood[1,2], Adam H. Steeves[1], and Heather J. Kulik[1,]*

[1]*Department of Chemical Engineering, Massachusetts Institute of Technology, Cambridge, MA 02139*
[1]*Department of Chemistry, Massachusetts Institute of Technology, Cambridge, MA 02139*

**\*Corresponding Author:** email: hjkulik@mit.edu; mail: Department of Chemical Engineering, 77 Massachusetts Ave Rm 66-464, Cambridge, MA 02139, phone: 617-253-4584

[#]**email addresses:** vyshnavi@mit.edu (V.V.); rimsha@mit.edu (R.M.); anazemi@mit.edu (A.N.); asteeves@gmail.com (A.H.S.)

ABSTRACT: Computational prediction of enzyme mechanism and protein function requires accurate physics-based models and suitable sampling. We discuss recent advances in large-scale quantum mechanical (QM) modeling of biochemical systems that have reduced the cost of high-accuracy models. Trade-offs between sampling and accuracy have motivated modeling with molecular mechanics (MM) in a multi-scale QM/MM or iterative approach. Limitations to both conventional density functional theory (DFT) and classical MM force fields remain for describing non-covalent interactions in comparison to experiment or wavefunction theory. Because predictions of enzyme action (i.e., electrostatics), free energy barriers, and mechanisms are sensitive to the protocol and embedding method in QM/MM, convergence tests and systematic methods for quantifying QM-level interactions are a needed, active area of development.

**Introduction**

Quantum mechanical (QM) modeling is indispensable for the prediction of enzymatic mechanisms and the accurate description of the energetics of bond rearrangement. Advances in computing power and algorithms[1-4] have made study of large systems (ca. 100–1000 atoms) tractable with first-principles density functional theory (DFT) modeling[5-7] (**Figure 1**). Nevertheless, most DFT functionals lack predictive accuracy in describing key properties most relevant to protein modeling such as non-covalent interactions[8,9] or metal-catalyzed reactions[10-12]. Complementary advances in reducing the scaling of correlated wavefunction theory (WFT), e.g. with local variants of coupled cluster[13,14], and in decomposing interactions[15,16], e.g. with symmetry-adapted perturbation theory (SAPT)[16,17], are shedding light on limitations[18-22] of DFT-level and lower models (**Figure 1**). At the same time, researchers continue to improve semi-empirical methods[23] and corrections to DFT[24,25] to achieve low-cost approximations with higher accuracy. This low cost is needed to sample reactive configurations and to discover intermediates rather than relying on static (e.g., experimental crystal) structures for mechanistic interpretations (**Figure 1**).

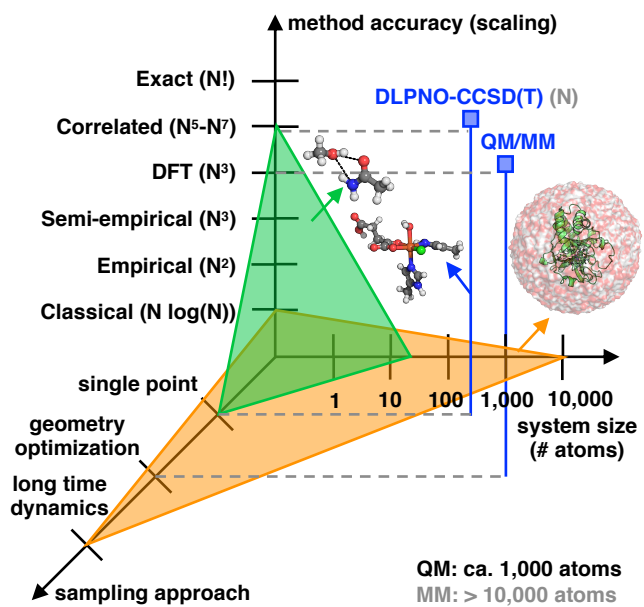

**Figure 1.** A depiction of the interplay between the system size (number of atoms in the system), level of theory (accuracy of the method and its scaling with system size), and sampling approach (i.e., single-point calculation, geometry optimization, or long-time dynamics) in simulations of biological systems. The traditional combination of trade-offs needed to study systems with highly accurate correlated wavefunction theory is represented by a green translucent region, illustrated by a system annotated by a green arrow. The traditional combination of level of theory and sampling applied to an entire protein is shown as an orange translucent region, illustrated by representative system annotated by an orange arrow. Modern methods that push outside the boundaries of these tradeoffs are shown as blue sticks: correlated wavefunction theory (i.e., DLPNO-CCSD(T) which can scale linearly with a large prefactor) and QM/MM (i.e., QM is DFT) with up to 1,000 atoms in the QM region during geometry optimizations.

Despite advances, modeling trade-offs remain that typically must be assessed by a researcher for each new system. Although the time scales over which purely classical simulation can be carried out have also increased[26], molecular mechanics (MM) force fields remain considerably less accurate[27] than QM methods and cannot readily sample reactive events (**Figure 1**). It is thus not possible to both exhaustively sample reaction mechanisms and protein motion while modeling large systems at a high level of theory (**Figure 1**). Some of the most pressing considerations include: i) the level and specific type of theory chosen for modeling; ii) the nature of embedding and size of system treated with high-level methods in multi-scale (i.e., QM/MM) modeling; iii) the degree and type of sampling approach (**Figure 1**). While researchers often rely on intuition to choose the reactive intermediates to study or level of theory to use, systematic approaches relying less on prior knowledge are increasingly favored for their ability to avoid bias. In this Opinion, we describe how researchers are leveraging advances in large-scale QM for fully *ab initio* and multi-scale QM/MM protein modeling, and we discuss how these approaches are being bridged by adaptive and systematic model selection tools to use QM only when it is most needed.

**Modeling proteins accurately with fast and large-scale QM**

Although large-scale QM (e.g., with DFT) is increasingly tractable for the study of entire proteins[5-7], its high computational cost has limited its use in comparison to MM (**Figure 1**). Where DFT has been applied with a reliable functional, it has provided[5,7] superior estimates of interactions relevant to disordered peptides or loop motifs that are common in protein active sites. Even as DFT gets faster and better functionals and corrections are developed, higher-scaling WFT methods are required to accurately describe the van der Waals packing and hydrogen bonding that determine enzyme structure and dynamics.

Numerous researchers have applied high-accuracy QM for screening and analysis of high-resolution X-ray crystal structures in the protein databank (PDB) for method benchmarking[18,20] or analysis[28-31] of non-covalent interactions. An example of a benchmark dataset is the BioFragment database[32], which contains 3,380 sidechain–sidechain and 100 backbone–backbone interactions extracted from high-resolution PDB structures and evaluated with correlated WFT (e.g., SAPT or coupled cluster theory, CCSD(T)). This and more recently developed high-level computed sets (e.g., PEPCONF[33] or the NCI Atlas[22]) have been used to validate semi-empirical theories, DFT functionals, or local approximations to coupled cluster theory.[7]

Large-scale analysis of the PDB has recently been leveraged to reveal the mechanistic importance of non-covalent interactions.[28-31] For example, both π-stacking[34,35] and cation–π interactions (e.g., in Lys/Arg–Tyr/Phe interactions)[28] were analyzed by comparing WFT (i.e., SAPT and DLPNO-CCSD(T)) favored geometries of model systems to structures observed in the PDB, revealing good correspondence of models for dispersive/electrostatic Arg–arene interactions, whereas a minimal model (i.e., $NH_3^+$) system failed to describe the strongly electrostatic Lys interaction.[28] Given that expectations of non-covalent distances influence the process of X-ray crystal structure solution, Qi et al.[30] mined a set of over 10,000 reasonable resolution (i.e., 2.0 Å or better) protein crystal structures to detect unexpectedly close non-covalent distances (i.e., < 85% of the sum of van der Waals radii of the shortest non-bonded heavy-atom distances). All high-resolution structures and most amino acid pairs were observed to have close contacts (**Figure 2**).[30] WFT (i.e., SAPT) confirmed that short non-covalent separations (ca. 2.5 Å) were favorable even between neutral amino acid sidechains.[30]

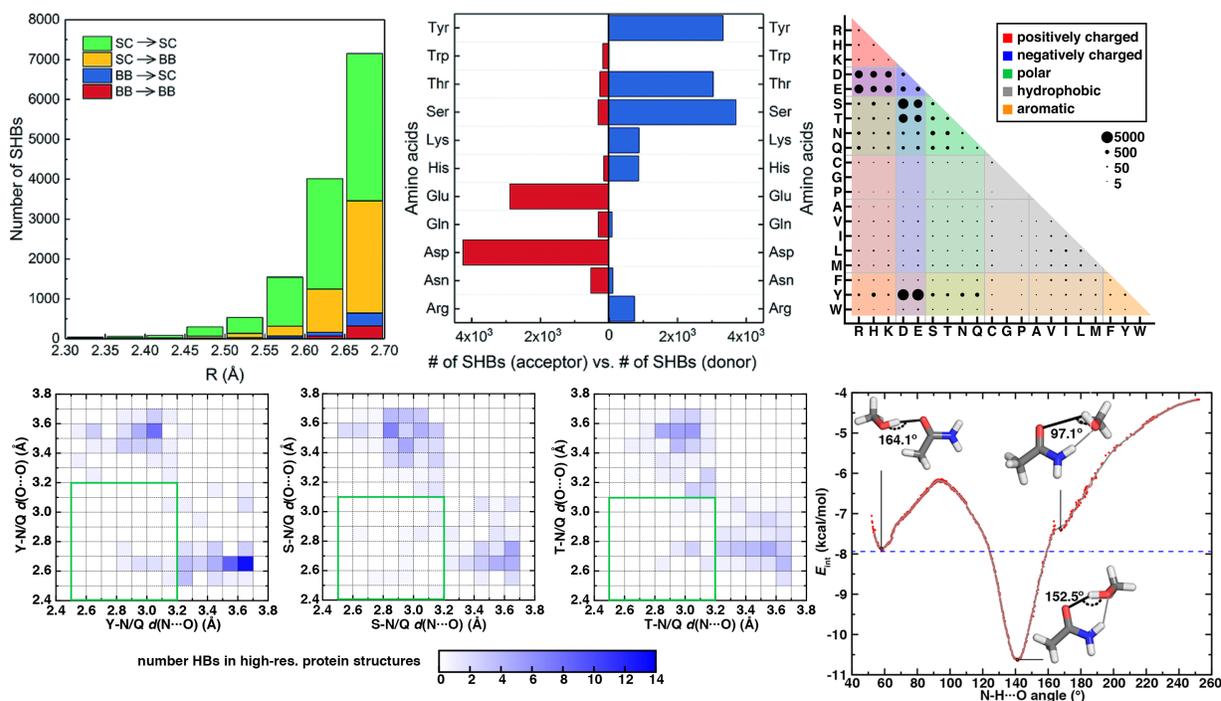

**Figure 2.** (top, left and middle) Analysis of hydrogen bonds (HBs) in 1,663 high-quality protein structures from Ref. [29]: stacked histogram (0.05 Å bin width) of shortest heavy-atom distances grouped by backbone (BB) or sidechain (SC) interaction type (left) and prevalence of donor or acceptor short hydrogen bond interactions (SHB, middle) over the same set. (top, right) Number of close contacts (< 85% van der Waals' separation) over 11,569 good quality protein structures from Ref. [30]. (bottom, left panes) Two-dimensional histogram with 0.1 Å bin widths (colored according to inset colorbar) of the shortest simultaneous donor and acceptor distances between Asn or Gln and Tyr (left), Ser (middle), and Thr (right) on a set of 3,908 residue pairs from high-quality protein structures in Ref. [31]. The green box in each indicates the presence of two HBs. (bottom, right) Reaction coordinate[31] for interconversion from an O–H⋯O HB to an ambifunctional HB and to an N–H⋯O HB in a model system of Ser/Thr–Asn/Gln with DLPNO-CCSD(T). Reproduced with permission from refs. [29], [30], and [31], published by The Royal Society of Chemistry 2019, the American Chemical Society 2019, and The Royal Society of Chemistry 2021, respectively.

Zhou and Wang studied[29] a higher-resolution subset of proteins and confirmed the prevalence of short hydrogen bonds (HBs, ca. 2.5–2.7 Å), contrary to expectations from conventional force fields[29,30] (**Figure 2**). Although Zhou and Wang showed that amino acids have a bias for forming either shorter donor or acceptor interactions[29], unusually short distances between protein residues that simultaneously act as ambifunctional HB donors and acceptors have recently been characterized (**Figure 2**).[30,31] Vennelakanti et al.[31] used high-accuracy references (i.e., DLPNO-CCSD(T)) to establish the expected[30,36] favorability of ambifunctional Ser/Thr/Tyr–Asn/Gln HBs (**Figure 2**). These ambifunctional HBs were observed in numerous PDB structures, supporting the WFT-predicted stabilization especially of aliphatic hydroxyl (i.e., Ser/Thr) ambifunctional interactions that were missed by force field modeling.[30,31,36] As high-accuracy methods become increasingly tractable, further studies will undoubtedly reveal previously poorly understood mechanistic roles for non-covalent interactions.

**Multi-scale methodology and QM/MM convergence**

Despite the well-recognized promise of multi-scale QM/MM modeling, its predictive accuracy is challenged by complexities and trade-offs in the choice of embedding approach (i.e., mechanical, electrostatic, or polarizable) or boundary treatment (i.e., charges and capping of cleaved bonds) as well as which subsystem to treat with QM, as discussed in a recent review.[37] Although polarizable embedding has been shown to be beneficial[38] in some cases (e.g., photo-excitations), its impact on some properties[39] (e.g., redox potentials[40]) is less clear and can introduce new theoretical challenges.[38]

Advances over the last decade in accelerated, large-scale QM have enabled systematic, radial convergence studies of properties in QM/MM. Here, the dependence of properties on QM region can be judged against large, 1000-atom or larger QM regions in QM/MM models or large QM clusters or carried out by comparison to the largest (typically ca. 600–700 atom) tractable QM region. Most such studies have revealed an exceptionally slow approach to the asymptotic limit of full QM treatment including for properties such as: non-covalent distances[41,42], NMR shieldings,[43,44] proton transfer,[45-47] solvation effects,[48] barrier heights or isomerization energies,[49-52] forces[53], partial charges[54,55] and charge transfer,[51] bond critical points[56], and redox[57,58] or electrostatic[55,59] potentials (**Figure 3**). When the QM region is naively selected based on distance alone, a wide range of properties of the geometric and electronic structure require between 400 and 800 atoms to achieve reasonably converged properties (**Figure 3**). Although there exist exceptions where the structure of an enzyme may limit the dependence on QM region, these exceptions are few and system-specific.[60,61]

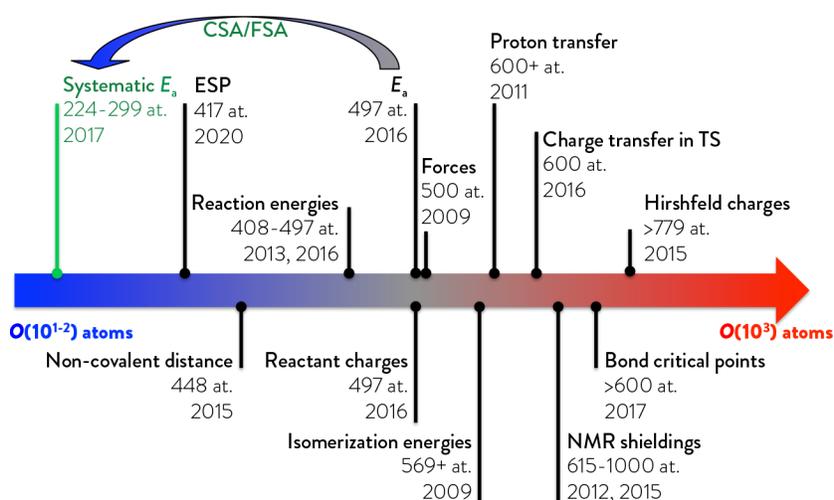

**Figure 3.** Schematic of radial (black line and text) and systematic (green line and text, selected with CSA or FSA[51,60], as indicated by curved arrow) convergence of properties with respect to asymptotic limits shown qualitatively ranging from small QM region sizes (10–100 atoms) to large QM region sizes (1000 atoms) in QM/MM calculations on a range of enzymes. Each property is annotated with the number of atoms needed and year of publication, as discussed and referenced in the main text.

**Configurational sampling challenges**

While large-scale QM electronic structure single-point energy evaluations are increasingly tractable, configurational sampling is also important for understanding dynamics relevant to enzyme catalysis. Recent studies have aimed to quantify the relative importance of these two considerations to treat them in a balanced fashion.[55,62] Sampling configurations from classical molecular dynamics (MD) and pairing them with QM/MM property evaluation is an efficient alternative to full-QM sampling, but this approach has been criticized when the MM-derived structures are poor approximations of relaxed QM structures.[62] Special concerns arise when the reaction coordinate is not known *a priori* or is sensitive to long-time dynamical rearrangements of the protein. Modest conformational barriers determined by subtle non-covalent interactions mean that the preferred positioning of sidechains or the substrate can differ unpredictably from those observed in a crystal structure and can change on long (ca. ns to μs) timescales, as we have shown in non-heme Fe(II) enzymes.[36,63] In such cases, MM dynamics for extensive sampling must be balanced with large-QM-region QM/MM geometry optimizations to quantify critical interactions[36] (e.g., covalent contributions to HBs[64]). We also recently showed[55] that while properties approach asymptotic limits slowly with increasing QM region size, the fluctuations of a property of a representative metal binding site of Zn-dependent DNA methyltransferase were as large as the differences between configurationally averaged values for different size QM regions. Nevertheless, these fluctuations could be well captured with a small QM region, motivating a strategy of using small QM regions for sampling and large QM regions for property evaluation.[55]

**Systematic and adaptive QM/MM**

Because intuition about an enzyme mechanism is often limited, systematic approaches to QM region selection are needed to both improve predictive accuracy in QM/MM and provide insight into the electronic environment of an enzyme active site. Early systematic methods that focused on computing the effect of MM region perturbations[49,65] (e.g., elimination of MM electrostatic contributions[49]) suffered from overestimation of the impact of charged residues on the QM region. As rapid, large-scale QM calculations have become increasingly tractable, new systematic methods for QM region construction have specifically targeted interactions that cannot be described at the MM level (e.g., charge transfer).[60] One such technique, charge shift analysis (CSA)[51,60] constructs a QM region from all residues that exhibit charge transfer (i.e., as judged by the by-residue summed partial charges) with a core active site.[60] The detected residues are then tested to ensure that only the most essential residues are included in the QM region, reducing converged QM regions by 2–3-fold from radial estimates (**Figure 3**).[60]

Conceptual DFT and density analysis have long contributed to our understanding of enzyme catalysis. For example, Fukui functions[66,67], spatial representations of the most reactive regions of a molecule from frontier orbital theory, have been used to identify the active site of enzymes.[68] In Fukui shift analysis (FSA), QM regions are constructed by identifying which residues alter the Fukui function of the active site residues (**Figure 3**).[60] Other contemporary examples include methods and software that propose the use[69,70] of bond critical points from the quantum theory of atoms in molecules, the electron localization function[71], or contact networks of interaction energies[57,72,73] to define appropriate QM regions. Recently, self-parameterizing hybrid methods were developed to automate convergence of central QM forces in QM/MM construction.[74] Conceptual DFT heuristics[75], although not yet demonstrated in QM/MM, have been exploited to guide discovery of reaction mechanisms in a way that could be beneficial in future multi-scale studies of enzymatic catalysis.

Techniques to use adaptive QM regions in QM/MM represent a complementary area of development.[76-78] Here, a moving indicator is combined with a buffer region that ensures smooth forces and dynamics as the QM region adjusts[76,78]. The higher computational cost of the included buffer region and the need for an indicator[79,80] has limited adaptive QM/MM primarily to proton transfer studies[79,80] with semi-empirical QM treatments.[81] While promising[76,77], further extension to diverse classes of enzymes and higher accuracy methods is needed. Improvements leveraging the descriptors from systematic QM region construction methods are expected to contribute to these developments.

*Ab initio* **and multi-scale molecular dynamics**

Given the challenges associated with simultaneous treatment of electronic structure and sampling, multi-scale QM/MM MD is typically employed with both small QM regions (e.g., 30–60 atoms[82-84]) and with low-cost, semi-local DFT[85] or semi-empirical QM[86-88] (e.g., PM6 or DFTB[89]). To carry out free energy simulations, umbrella sampling has been used by multiple groups[86,87,90] along the methyl transfer coordinate of catechol *O*-methyltransferase (COMT, **Figure 4**). Across all studies[86,87,90], significant (ca. 6–10 kcal/mol) changes in free energy barriers were observed with changes in QM region or level of QM theory (**Figure 4**). In the smallest QM regions, the significant shift from charge-separated reactants to charge-neutralized products is unfavorable, leading to prediction by all methods of an endothermic reaction unless larger QM regions are used (**Figure 4**). Notably, for some lower-cost, semi-empirical methods, a larger QM region does not necessarily lead to a better agreement with the experimental free energy barrier (ca. 18–19 kcal/mol), which is instead underestimated (**Figure 4**). Recent comparisons[83] of semi-empirical and hybrid DFT free energy profiles have confirmed that semi-empirical methods fail to predict accurate free energies for enzymatic systems.

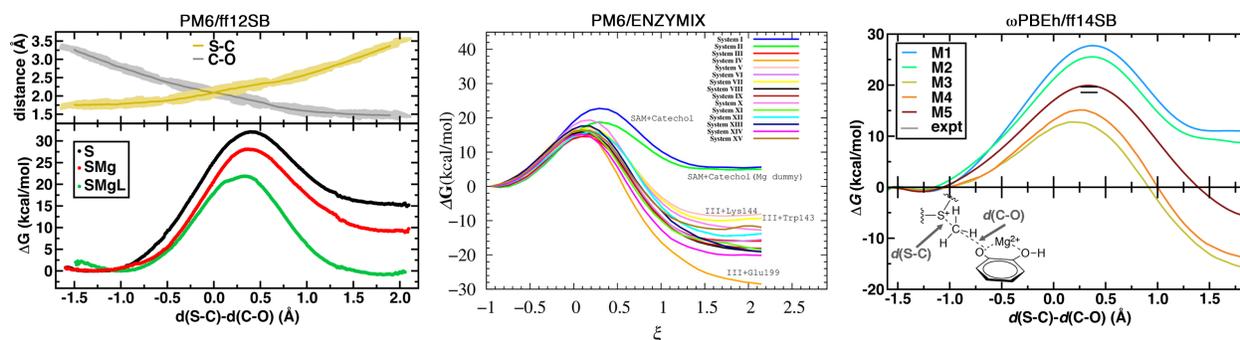

**Figure 4.** QM/MM free energy simulations of the barrier in the enzyme catechol *O*-methyltransferase using a linear combination of distance differences coordinate for the transfer of a methyl group to the catecholate substrate from three independent studies: (left) PM6/ff12SB, (middle) PM6/ENZYMIX, and (right) ωPBEh/ff14SB. QM region definitions that are most comparable across each study are as follows: (left) S: substrates only, SMg: substrates and $Mg^{2+}$, SMgL: substrate, $Mg^{2+}$ and its coordination sphere, and Lys144; (middle) SAM+catechol: substrates only, SAM+Catechol (Mg dummy): substrates and $Mg^{2+}$ dummy atom, III+Lys144: substrate, $Mg^{2+}$ and its coordination sphere sidechains, and Lys144 sidechains; (right) M1: substrates and $Mg^{2+}$, M2: substrate, $Mg^{2+}$ and its coordination sphere, M3 and M4: 170- and 325-atom QM region calculations, and M5: a 544-atom radially-converged QM region. Reproduced with permission from refs. [86], [87], and [90], published by PLoS 2016, the American Chemical Society 2016, and PCCP Owner Societies 2018, respectively.

As DFT-level QM/MM free energy simulations become more tractable, researchers will need to balance suitable sampling with appropriate levels of theory and QM region sizes to obtain robust and reproducible QM/MM free energy profiles. Systematic QM/MM methods that worked well for static calculations exhibit higher variability in free energy simulations due to greater differences in flexibility with change in QM region (**Figure 4**). Leveraging GPU-accelerated DFT, fully *ab initio* MD was very recently carried out with a range-separated hybrid functional[91] to evaluate the length-scales over which QM properties are coupled in proteins. Significant, long-range interactions were observed in pairs of residues that included at least one mobile, polar or charged residue. Consistent with observations from cross-correlation analysis in QM/MM MD on COMT[90,92], transient, non-covalent interactions result in significantly stronger coupling than is observed for shorter-range, covalently bound residue pairs (e.g., disulfide bridges[91] or $Mg^{2+}$-carboxylate bonds[90,92]). Beyond small systems, analysis of charge coupling is not yet broadly tractable without resorting to lower-level theory (e.g., to semi-empirical methods[89]) but could become a useful tool to provide guidance on how to select systematic or adaptive QM regions for robust QM/MM free energy simulations.

**Conclusions**

Faster computational chemistry now makes it possible to study enzymes with better fidelity than ever before. Nevertheless, the dynamic nature of proteins and the large number of ways the enzyme environment can catalyze bond rearrangement require a degree of sampling that still challenges even low-cost, molecular mechanics and semi-empirical techniques. This tension has historically been resolved by compromises selected with guidance from chemical intuition and experimental data. As discussed in this Opinion, a number of systematic studies have challenged the conclusions reached with more approximate methods. Approaches that iterate between classical and first-principles modeling have also been essential for uncovering non-covalent interactions not expected from intuition. New developments in the accuracy and efficiency of low-[89] to mid-cost[1-4] physics-based models will strengthen the role of high-accuracy modeling in enzyme catalysis studies. Machine learning will continue to play an increasing role in biological modeling.[93] Such advances will not be confined to further improvements to machine learning potentials[94] but will also involve systematic, statistical models to reveal modes of enzyme action[91,95,96] and to guide single- or multi-scale method selection[60,74,97,98], reducing the barriers to mechanistic study in the absence of *a priori* knowledge.

DECLARATION OF INTEREST

The authors declare no conflict of interest.

ACKNOWLEDGMENT

This work was supported by the National Science Foundation grant numbers CBET-1704266 and CBET-1846426; the Burroughs Wellcome Fund Career Award at the Scientific Interface Grant 1010755; and Exxon Mobil Research and Engineering.